\documentclass[conference]{IEEEtran}
\IEEEoverridecommandlockouts
\usepackage{tex/packages}
\usepackage{flushend}
\def\BibTeX{{\rm B\kern-.05em{\sc i\kern-.025em b}\kern-.08em
    T\kern-.1667em\lower.7ex\hbox{E}\kern-.125emX}}
\begin{document}

\title{The Continuous-Time RC-Chain ADC}

\author{\IEEEauthorblockN{Hampus Malmberg}
\IEEEauthorblockA{\textit{Dept. of Information Technology \& Electrical Engineering} \\
\textit{ETH Zürich}\\
Zürich, Switzerland \\
malmberg@isi.ee.ethz.ch}
\and
\IEEEauthorblockN{Fredrik Feyling}
\IEEEauthorblockA{\textit{Dept. of Electronic Systems} \\
\textit{Norwegian University of Science and Technology (NTNU)}\\
Trondheim, Norway\\
fredrik.e.feyling@ntnu.no}
}

\IEEEpubid{This work has been submitted to the IEEE for possible publication. Copyright may be transferred without notice, after which this version may no longer be accessible.}

\maketitle

\begin{abstract}
An amplifier-less continuous-time analog-to-digital converter consisting of only passives, comparators, and inverters is presented. Beyond simplicity, the architecture displays significant robustness properties with respect to component variations and comparator input offsets. We give an analytical design procedure demonstrating how to parameterize the architecture to a range of signal-to-noise and bandwidth requirements and validate the procedure's accuracy with behavioral transient simulations.

\end{abstract}

\begin{IEEEkeywords}
\Ac{ADC}, continuous-time ADC, amplifier-less ADC, oversampling ADC, control-bounded ADC.
\end{IEEEkeywords}

\section{Introduction}

In recent years, \acp{CTADC} have gained increasing popularity due to their inherent anti-alias filter and reduced driving requirements.
Compared to a signal chain consisting of an anti-alias filter, a driving buffer and a \ac{DTADC}, a \ac{CTADC} is often superior in terms of area and power consumption \cite{S:23,PMB:23}. 
The dominating \ac{CTADC} architectures are the (single-stage or cascaded) \ac{CTSDM} and the \ac{CTP} \ac{ADC}. 

Common to these \acp{CTADC} is the use of operational amplifiers or transconductors to realize internal continuous-time filters.
Due to demanding noise and linearity requirements, amplifiers are often the single most power-hungry component for \ac{CTADC} \cite{R:15}. 
Furthermore, the use of operational amplifiers requires high supply voltages and does not benefit from technology scaling \cite{SCY:08}.

To reduce the number of amplifiers, several hybrid active/passive \acp{CTSDM} have been presented \cite{SCY:08, D:05, NPG:16}. By replacing some of the active elements with passive RC-networks, higher-order noise shaping can be achieved with fewer amplifiers. Although this technique may reduce the power consumption of the loop filter, technology and supply voltage must still be chosen with the amplifier design in mind. While some fully-passive \ac{DT}-\acsp{SDM} have been presented \cite{CL:97,YQA:14}, to the best of the authors' knowledge, no amplifier-less \ac{CTADC} has ever been published.

In this paper, we seek to combine the benefits of inherent anti-aliasing filtering, resistive input impedance, and amplifier-less design; to this end, we propose the RC-chain \ac{ADC}, shown in \Fig{fig:RC-chain}, a first of its kind, amplifier-less \ac{CTADC}. 
The system consists of a chain of RC filters combined with single-bit digital feedback loops. 
Its operation revolves around reducing the voltage swing over the last capacitor in the chain, using local digital feedback, given an input signal $v_u$. 

The purpose of this paper is to provide a pre-study of the RC-chain that lays the foundations for a future \ac{IC} implementation; as such, we rely on mathematical analysis and behavioral transient simulations to assess the  robustness, feasibility, and sizing requirements of a future prototype.

The paper is structured as follows:
\Sec{sec:rc_filter} presents the general operating principle of the RC-chain \ac{ADC}. 
\Sec{sec:estimation} covers the accompanying digital reconstruction filtering, \Sec{sec:expected_performance} gives a design procedure that dimensions the \ac{ADC} to meet a target \ac{SNR} and \ac{BW} requirement, and \Sec{sec:behavioral_simulations} validates the design approach with behavioral simulations. \Sec{sec:example} assesses sensitivity to component variations, digital loop delay, gives thermal noise estimates, and comments on the input impedance of the \ac{ADC}, all in the context of a nominal $10$-bit, $10 \, \si{\mega \hertz}$ bandwidth \ac{ADC} example. \Sec{sec:outlook} summarizes the potential strengths and pitfalls of the proposed architecture and finally \Sec{sec:conclusion} concludes our findings.

\IEEEpubidadjcol
\section{Operating Principle}\label{sec:rc_filter}
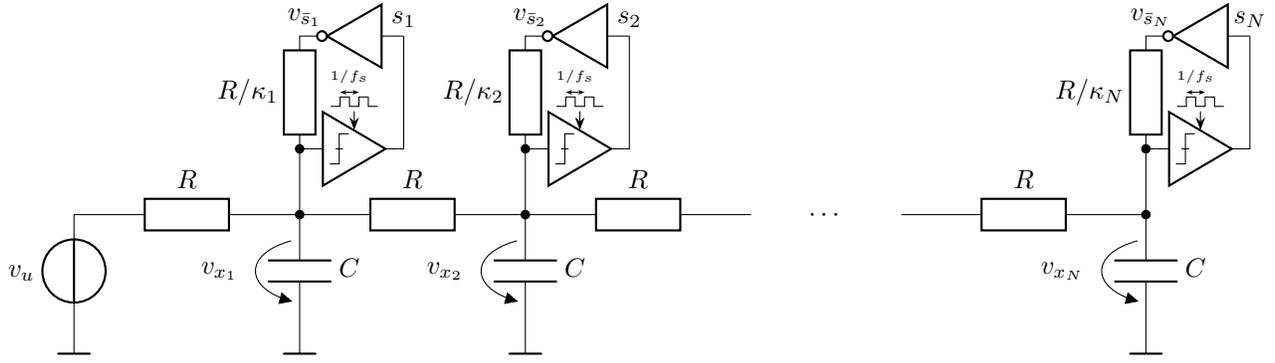
\begin{figure*}[tbp]
    \begin{center}
            \begin{circuitikz}[european]

    \draw 
    (0,0) 
    node[rground] {} 
    to[V,l=$v_u$] (0, 1.5)
    to[R,l=$R$,-*] (3, 1.5)
    coordinate (x1) {}
    to[R,l=$R$,-*] (6, 1.5)
    coordinate (x2) {}
    to[R,l=$R$,] (9, 1.5)
    coordinate (x3) {};
    \draw 
    (10, 1.5)
    node {$\cdots$}
    ++(1, 0)
    to[R,l=$R$,-*] (14.25, 1.5)
    coordinate (xN) {}
    to[C,l=$C$,v=$v_{x_N}$] (14.25, 0)
    node[rground] {};

    \draw (x1) to[C,l=$C$,v=$v_{x_1}$] ++(0,-1.5) node[rground] {};
    \draw (x2) to[C,l=$C$,v=$v_{x_2}$] ++(0,-1.5) node[rground] {};

    \coordinate (x_upper) at ($(x1) + (0, 0.875)$) {};
    \draw
    ($(x_upper) + (0.725, 0)$) node[plain mono amp, scale=0.5] (comp) {}
    ++(0,1.5) node[ieeestd not port, scale=0.75, xscale=-1] (inv) {}
    (inv.out) to[short] ($(x_upper) + (0, 1.5)$)
    to[R,l_=$R / \kappa_1$,-*] (x_upper)
    (x_upper) to[short] (comp.in)
    (comp.out) -| (inv.in)
    ($(comp) + (-0.325, -0.2)$) -- ++ (0.125,0)
    -- ++(0,0.4)
    -- ++(0.125,0)
    ($(comp) + (-0.25, 0)$) to[thick] ++(0.1,0)
    ($(comp) + (0, 0.5)$) coordinate (clk)
    ($(clk) + (-0.3125,0.065)$) 
    -- ++(0.125, 0)
    -- ++(0, 0.125)
    coordinate (clks)
    -- ++(0.125, 0)
    -- ++ (0, -0.125)
    -- ++(0.125, 0)
    -- ++(0, 0.125)
    coordinate (clke)
    -- ++(0.125, 0)
    -- ++(0, -0.125)
    -- ++(0.125, 0)
    (inv.in) node[above] {$s_1$}
    (inv.out) node[above] {$v_{\bar{s}_1}$}
    (x_upper) to[short] (x1)
    ;
    \draw[{Stealth[length=0.875mm]}-{Stealth[length=0.875mm]}] ($(clks) + (0, 0.0625)$) -- node[above] {\tiny$1/f_s$} ($(clke) + (0, 0.0625)$);
    \draw[-{Stealth}] (clk) -- ++(0,-0.25);

    \coordinate (x_upper) at ($(x2) + (0, 0.875)$) {};
    \draw
    ($(x_upper) + (0.725, 0)$) node[plain mono amp, scale=0.5] (comp) {}
    ++(0,1.5) node[ieeestd not port, scale=0.75, xscale=-1] (inv) {}
    (inv.out) to[short] ($(x_upper) + (0, 1.5)$)
    to[R,l_=$R / \kappa_2$,-*] (x_upper)
    (x_upper) to[short] (comp.in)
    (comp.out) -| (inv.in)
    ($(comp) + (-0.325, -0.2)$) -- ++ (0.125,0)
    -- ++(0,0.4)
    -- ++(0.125,0)
    ($(comp) + (-0.25, 0)$) to[thick] ++(0.1,0)
    ($(comp) + (0, 0.5)$) coordinate (clk)
    ($(clk) + (-0.3125,0.065)$) 
    -- ++(0.125, 0)
    -- ++(0, 0.125)
    coordinate (clks)
    -- ++(0.125, 0)
    -- ++ (0, -0.125)
    -- ++(0.125, 0)
    -- ++(0, 0.125)
    coordinate (clke)
    -- ++(0.125, 0)
    -- ++(0, -0.125)
    -- ++(0.125, 0)
    (inv.in) node[above] {$s_2$}
    (inv.out) node[above] {$v_{\bar{s}_2}$}
    (x_upper) to[short] (x2)
    ;
    \draw[{Stealth[length=0.875mm]}-{Stealth[length=0.875mm]}] ($(clks) + (0, 0.0625)$) -- node[above] {\tiny$1/f_s$} ($(clke) + (0, 0.0625)$);
    \draw[-{Stealth}] (clk) -- ++(0,-0.25);

    \coordinate (x_upper) at ($(xN) + (0, 0.875)$) {};
    \draw
    ($(x_upper) + (0.725, 0)$) node[plain mono amp, scale=0.5] (comp) {}
    ++(0,1.5) node[ieeestd not port, scale=0.75, xscale=-1] (inv) {}
    (inv.out) to[short] ($(x_upper) + (0, 1.5)$)
    to[R,l_=$R/\kappa_N $,-*] (x_upper)
    (x_upper) to[short] (comp.in)
    (comp.out) -| (inv.in)
    ($(comp) + (-0.325, -0.2)$) -- ++ (0.125,0)
    -- ++(0,0.4)
    -- ++(0.125,0)
    ($(comp) + (-0.25, 0)$) to[thick] ++(0.1,0)
    ($(comp) + (0, 0.5)$) coordinate (clk)
    ($(clk) + (-0.3125,0.065)$) 
    -- ++(0.125, 0)
    -- ++(0, 0.125)
    coordinate (clks)
    -- ++(0.125, 0)
    -- ++ (0, -0.125)
    -- ++(0.125, 0)
    -- ++(0, 0.125)
    coordinate (clke)
    -- ++(0.125, 0)
    -- ++(0, -0.125)
    -- ++(0.125, 0)
    (inv.in) node[above] {$s_N$}
    (inv.out) node[above] {$v_{\bar{s}_N}$}
    (x_upper) to[short] (xN)
    ;
    \draw[{Stealth[length=0.875mm]}-{Stealth[length=0.875mm]}] ($(clks) + (0, 0.0625)$) -- node[above] {\tiny$1/f_s$} ($(clke) + (0, 0.0625)$);
    \draw[-{Stealth}] (clk) -- ++(0,-0.25);

\end{circuitikz}
        \caption{\label{fig:RC-chain}The RC-chain \ac{ADC} where for $1 > \kappa_1 > \kappa_2 > \dots > \kappa_N$ the clocked digital feedback loops successively reduce the voltage swing over the capacitances such that $\max |v_u| > \max |v_{x_1}| > \max |v_{x_2}| > \dots > \max |v_{x_N}|$ and where the resulting \ac{SNR} is inversely proportional to $\max |v_{x_N}|$. }
    \end{center}
\end{figure*}

At first glance, the experienced \ac{ADC} designer will be tempted to interpret the bitstreams $s_1,\dots,s_N$, in \Fig{fig:RC-chain}, as filtered, sampled, and quantized versions of the input $v_u$. 
In our experience, this perspective will make the analysis of \Fig{fig:RC-chain} difficult. 
Instead, we encourage the reader to momentarily forget the comparators in \Fig{fig:RC-chain} and acknowledge that if $\max |v_{x_N}|$ is ``small'' the net effort of $v_{\bar{s}_1}, \dots, v_{\bar{s}_N}$, seen over time, must partially cancel $v_u$ as it passes through the RC-chain filter.
Furthermore, since $v_u, v_{\bar{s}_1}, \dots, v_{\bar{s}_N}$ all add linearly to the circuit, there must be a linear mapping from the sequences $s_1, \dots, s_N$ that approximates a filtered version of $v_u$ with an accuracy that increases as $\max |v_{x_N}|$ decreases. 
In hindsight, this applies regardless of how $s_1, \dots, s_N$ comes about; however, the use of clocked comparators is practical when ensuring that $\max |v_{x_N}|$ remains ``small''.
This is the \ac{CBADC} perspective \cite{LBWB:11,LW:15,M:20,MWL:21,MFR:24} applied to the RC-chain \ac{ADC} in \Fig{fig:RC-chain}. What sets the RC-chain \ac{ADC} apart from prior \acp{CBADC} is the absence of linear gain in the analog filter.

\subsection{The ADC Output}\label{sec:estimation}
The final output of the RC-chain \ac{ADC} is neither $v_{x_N}$ nor $s_1,\dots,s_N$, instead, the estimated samples of $v_{u}$, follows
from 
\begin{IEEEeqnarray}{rCl}%
    \hat{u}[k] & = & \sum_{\ell=1}^N (h_\ell \ast s_\ell)[k] \label{eq:estimate}%
\end{IEEEeqnarray}
where $\ast$ refers to discrete-time convolution and the filters $h_1$, \dots, $h_N$ are learned through adaptive filtering as outlined in \cite{MMBFL:22, MFR:24}.
In the interest of space, we will not reiterate the details of finding $h_1$, \dots, $h_N$ 
but emphasize that the adaptive filtering step is a standard data-driven signal processing problem.
In terms of digital complexity, we restrict $h_1,\dots,h_N$ to be $32$-tap \ac{FIR} filters, where \Eq{eq:estimate} is operated at the Nyquist rate (typically much lower than $f_s$) and $s_1, \dots, s_N$ are decimated accordingly.

\subsection{Expected Performance}\label{sec:expected_performance}
By design, the RC-chain \ac{ADC} is inherently limited by the comparators sensitivity requirements; especially the rightmost comparator in \Fig{fig:RC-chain}
which will act on the smallest voltage swing $v_{x_N}$. 
We aim for a reduction in voltage swings of $\delta$ from one capacitor to the next, i.e., $|v_{x_\ell}| \leq \delta^\ell v_\text{max}$ where $v_\text{max} \eqdef \max |v_{u}|$.
By adapting the \ac{SNR} analysis in \cite{FMWLY:2023} to the RC-chain circuit in \Fig{fig:RC-chain}, we arrive at the approximation
\begin{IEEEeqnarray}{rCl}%
    \operatorname{SNR} &\approx & \epsilon \cdot \left(\frac{\delta^{2N}}{\omega_\mathcal{B}} \int_0^{\omega_\mathcal{B}}\frac{1}{|G(\omega)|^2} \dd \omega  \right)^{-1} \label{eq:expected_snr} 
\end{IEEEeqnarray}
where 
\begin{IEEEeqnarray}{rCl}%
    |G(\omega)| 
    & = & \left|\det\begin{pmatrix}
         \frac{\jmath \omega \tau + \xi_1}{\tau} & - \frac{1}{\tau} \\
        -\frac{1}{\tau} & \ddots & \ddots \\ 
        & \ddots & \frac{\jmath \omega \tau + \xi_{N-1}}{\tau} & -\frac{1}{\tau} \\
        && -\frac{1}{\tau} & \frac{\jmath \omega \tau + \xi_N}{\tau}
    \end{pmatrix}\right|^{-1} \nonumber \\ \label{eq:amplitude_response}
\end{IEEEeqnarray}
is the amplitude response from $v_u$ to $v_{x_N}$ when all comparators are inactive, 
$\det(\cdot)$ denotes the matrix determinant, $\omega_\mathcal{B}$ is the angular bandwidth of interest, $\xi_\ell \eqdef \kappa_\ell + 2$ except $\xi_N\eqdef \kappa_N + 1$, $\tau \eqdef RC$ is the time-constant, and $\epsilon$ is a proportionality constant related to the statistics of $v_{x_N}$.
The accuracy of \Eq{eq:expected_snr} rests on the assumption that $v_{x_N}$ is a white Gaussian noise process independent of $v_u$. For our purposes $\epsilon = \num{1.22e-3}$.

\subsubsection{Reduced Voltage Swings}\label{sec:ensuring_reduced_voltages}
We proceed by deriving the conditions for $\kappa_1$, \dots, $\kappa_N$ and $\tau$ that ensure $|v_{x_\ell}| \leq \delta^\ell v_\text{max}$. 
Considering the $\ell$-th RC-chain node in \Fig{fig:RC-chain}, 
the per node reduction $\delta$ can be sustained by considering two extreme cases: firstly, assume all capacitors are charged to their corresponding maximum allowed voltages as $\{v_{x_{\ell-1}}, v_{x_\ell}, v_{x_\ell+1}\} = \delta^{\ell - 1}\{1,\delta, \delta^2\} v_\text{max}$. In this scenario, we require that the digital feedback loop, i.e., $v_{\bar{s}_\ell}$ resist any further increase of $v_{x_\ell}$ by injecting an opposing current through $R/\kappa_\ell$. For $|v_{\bar{s}_\ell}| = v_\text{max}$,
\begin{IEEEeqnarray}{rCl}%
    \kappa_\ell & = & \delta^{\ell - 1}\left(1 + \delta ^2 - \delta\right) \label{eq:kappa_condition}
\end{IEEEeqnarray}
satisfy this requirement.
Secondly, we set the control rate $f_s$ such that $|v_{x_\ell}(t)|\leq \delta^\ell v_\text{max}$ for all $t\in[0,1/f_s)$. The worst possible combination of inputs occurs for $\{v_{x_{\ell-1}}, v_{s_\ell}, v_{x_\ell+1}\} = \{\delta^{\ell - 1}, \kappa_\ell, \delta^{\ell + 1}\} v_\text{max}$
which gives
\begin{IEEEeqnarray}{rCl}%
    |v_{x_\ell}(t)|
    & \leq & \frac{t}{\tau} \delta^{\ell-1}\left(2 (1 + \delta^2) - \delta\right) v_\text{max}.  \label{eq:largest_state}
\end{IEEEeqnarray}
Evaluating \Eq{eq:largest_state} at $t=1/f_s$ and equation with the upper bound $|v_{x_\ell}| = \delta^\ell v_\text{max}$, 
introducing the oversampling ratio $(\text{OSR})\eqdef f_s \pi / \omega_{\mathcal{B}}$, and finally rearranging yields
\begin{IEEEeqnarray}{rCl}%
    \tau & = & \frac{\pi (2(1 + \delta^2) - \delta) }{\delta \omega_\mathcal{B} \text{OSR}} \label{eq:tau}.
\end{IEEEeqnarray}

In summary, \Eq{eq:kappa_condition} and \Eq{eq:tau} ensure that the RC-chain's $N$-th capacitor will have a voltage swing $|v_{x_N}| \leq \delta^N v_\text{max}$ for any $\{ \delta, \omega_\mathcal{B}, \text{OSR}\}$ configuration.

\subsubsection{High-Level Design Procedure}\label{sec:high-level_design_procedure}
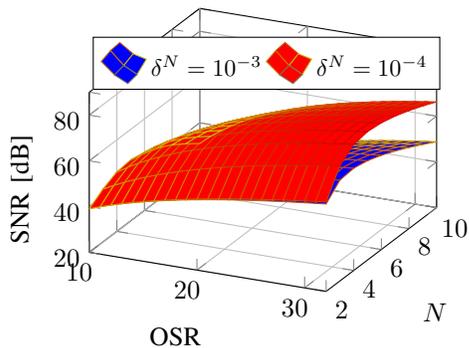
\begin{figure}
\centering
\begin{tikzpicture}
    \begin{axis}[
        xlabel={OSR},
        ylabel={$N$},
        zlabel={SNR [dB]},
        grid=major,
        grid=both,
        ytick={2,4,6,8,10},
        ztick={20,40,60,80},
        xmin=10,
        zmin=20,
        zmax=90,
        legend style={
            at={(0.01,0.9)}, 
            anchor=north west,
            legend columns=2,
            opacity=0.8,
            nodes={scale=0.9}, 
            },
        width=0.7\columnwidth,
        ]
        \addplot3[surf, color=blue, mesh/cols=9, opacity=0.99] file {data/expected_snr_0.001.txt};
        \addlegendentry{$\delta^N=10^{-3}$};
        
        \addplot3[surf, mesh/cols=9, color=red, opacity=0.4] file {data/expected_snr_0.0001.txt};
        \addlegendentry{$\delta^N=10^{-4}$};
    \end{axis}
\end{tikzpicture}
\caption{\label{fig:snr_surface}Expected \ac{SNR} \Eq{eq:expected_snr}, as a function of $N$ and $\text{OSR}$.}
\end{figure}

Applying \Eq{eq:kappa_condition} and \Eq{eq:tau} to \Eq{eq:expected_snr} and committing to an angular bandwidth $\omega_\mathcal{B}$, and a comparator design that is capable of sensing voltages of the order of $\delta^N v_\text{max}$ reduces the dependent variables of \Eq{eq:expected_snr} to $N$ and $\text{OSR}$.
Although analytical expression remains involved, due to \Eq{eq:amplitude_response}, the surface plot, shown in \Fig{fig:snr_surface}, appears smooth and concave with respect to both $N$ and \text{OSR}.
Furthermore, the \ac{SNR} generally increases with a larger $\text{OSR}$ and
has a local optimum in terms of $N$ dependent on both $\delta^N$ and $\text{OSR}$.

Finding a $\{N, \text{OSR}\}$ configuration for a target \ac{SNR}, $\omega_\mathcal{B}$ and $\delta^N$ can be done by searching the configuration space and evaluating \Eq{eq:expected_snr} exhaustively within a predetermined parameter range.
In terms of uniqueness, among all plausible solutions $\{N, \text{OSR}\}$, we favor the lowest $\text{OSR}$, as this reduces the speed requirements of the comparators. 

\subsection{Behavioral Simulations}\label{sec:behavioral_simulations}

To verify the accuracy of \Eq{eq:expected_snr}, each configuration from \Fig{fig:snr_surface}, i.e., $\text{OSR} \times N \times \delta^N = \{10,\dots,32\}\times\{2,\dots,10\}\times\{10^{-3},10^{-4}\}$, is validated by transient behavioral simulation.
For each configuration, two, $2^{12} \cdot \text{OSR}$ clock cycles long, simulations were performed. 
In the first simulation, $v_u$ takes piecewise constant values uniformly distributed over $[-v_\text{max}, v_\text{max}]$. 
In the second simulation $v_u$ is a sinusoidal signal of amplitude $v_\text{max}$ and angular frequency $\omega_{\mathcal{B}}/2$. 
Subsequently, the first simulation was used to train the adaptive filter, see \Sec{sec:estimation}, and the second simulation was used to measure \ac{SNR} directly from the \ac{PSD} of \Eq{eq:estimate}. 
From these simulations, we conclude that the expected \acp{SNR} \Eq{eq:expected_snr} agrees within $\pm2 \text{dB}$ of the simulated \acp{SNR} over the $352$ configurations in \Fig{fig:snr_surface}. 
Furthermore, the event $|v_{x_N}|= \delta^N v_\text{max}$ stands out as improbable, as we consistently measure root mean square values of $\operatorname{RMS}(v_{x_N} / v_\text{max} )  \approx \{\num{3.3e-4}, \num{3.4e-5}\}$ for $\delta^N = \{10^{-3}, 10^{-4}\}$ respectively.
 
\section{A Forth Order RC-Chain ADC}\label{sec:example}
Next we put \Sec{sec:high-level_design_procedure} to the test and seek an RC-chain design with a nominal $10$-bit resolution for a bandwidth $\omega_{\mathcal{B}} / (2 \pi) = 10$ \si{\mega \hertz} and comparators working with signal swings in the order of $10^{-3} v_\text{max}$. The proposed procedure results in $\text{OSR}=26$, $N=4$, $\tau \eqdef RC\approx 20$ \si{\nano \second}, and $\{\kappa_1,\dots,\kappa_4\}\approx\{\num{8.5e-01},\num{1.5e-01},\num{2.7e-02}, \num{4.8e-03}\}$.

\begin{figure}
    \centering
    \begin{tikzpicture}
        \begin{semilogxaxis}[
            xlabel={$\omega / \omega_\mathcal{B}$},
            ylabel={PSD [dBFS]},
            xmin=1e-3,
            xmax=2e0,
            ymin=-130,
            ymax=6,
            grid=both,
            height=0.35\columnwidth,
            width=\columnwidth,
        ]
        \addplot[black, thick] table[x=f_nom, y expr=\thisrow{psd}+42.27875, col sep=comma] {data/u_hat_psd.csv};

        \end{semilogxaxis}
    \end{tikzpicture}
    \caption{\label{fig:psd}\ac{PSD} of $\hat{u}[k]$, as in \Eq{eq:estimate}, from a $2^{12} \cdot \text{OSR}$ clock cycles long, behavioral simulations of a RC-chain \ac{ADC}, $N=4$, $\text{OSR}=26$, $\delta^N=10^{-3}v_\text{max}$. The y-axis is normalized to decibels relative to a full-scale input.}
\end{figure}
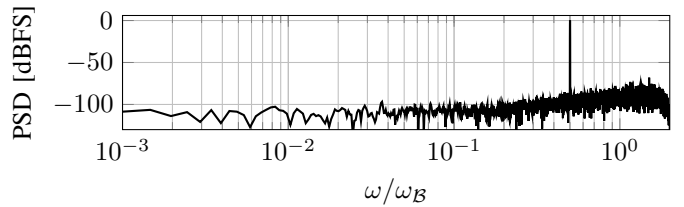

\begin{figure}
    \centering
    \begin{tikzpicture}
        \begin{axis}[
        ybar,
        xlabel={$v_{x_\ell} / (\delta^\ell v_\text{max})$},
        ylabel={Count},
        grid=both,
        height=0.375\columnwidth,
        width=\columnwidth,
        legend style={at={(0.0,1)}, anchor=north west, opacity=0.8},
        xtick={-1000,-500, 0, 500, 1000},
        xticklabels={$-1$, $-0.5$, 0, $0.5$, $1$},
        xmin=-1e3,
        xmax=1e3,
        ymin=0,
    ]
    \addplot[bar width=5378.95312876e-3/0.177827941, color=black, fill=black, opacity=1.0] table [x expr=\thisrow{b_1}/0.177827941*1e-3, y=c_1, col sep=comma] {data/state_test_hist.csv};
    \addlegendentry{$v_{x_1}$};
    \addplot[bar width=910.15635055e-3/0.0316227766, color=red, fill=red, opacity=0.7] table [x expr=\thisrow{b_2}/0.0316227766*1e-3, y=c_2, col sep=comma] {data/state_test_hist.csv};
    \addlegendentry{$v_{x_2}$};
    \addplot[bar width=143.80689093e-3/0.005623413252, color=blue, fill=blue, opacity=0.6] table [x expr=\thisrow{b_3}/0.005623413252*1e-3, y=c_3, col sep=comma] {data/state_test_hist.csv};
     \addlegendentry{$v_{x_3}$};
    \addplot[bar width=25.39887264e-3/1e-3, color=olive, fill=olive, opacity=0.7] table [x expr=\thisrow{b_4}*1e-3/1e-3, y=c_4, col sep=comma] {data/state_test_hist.csv};
    \addlegendentry{$v_{x_4}$};
     
    \end{axis}
    \end{tikzpicture}
    \caption{\label{fig:x_hist}Histograms of the voltages $v_{x_1}, \dots, v_{x_4}$, sampled at $k/f_s$ over $2^{12} \cdot \text{OSR}$ clock cycles, for a full swing sinusoidal input signal $v_u$ of angular frequency $\omega_\mathcal{B}  / 2$. Evidently, all voltage swings are well within their corresponding $\{\delta, \delta^2, \delta^3, \delta^4\}\approx\{\num{1.78e-1}, \num{3.16e-2}, \num{5.62e-3}, 10^{-3}\}$ bounds as per \Sec{sec:ensuring_reduced_voltages}.}
\end{figure}
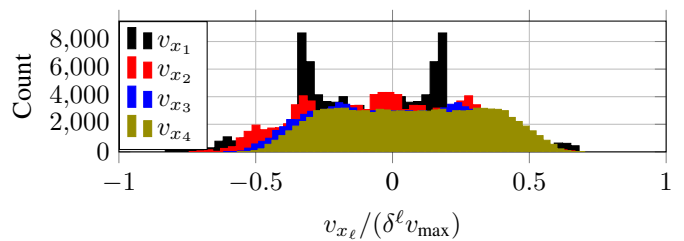

\Fig{fig:psd} shows the \ac{PSD} of \Eq{eq:estimate} from the transient simulation where the sinusoidal test signal rises out of the noise floor measuring a $\text{SNR}\approx 61.9 \text{dB}$, i.e., approximately $10.0$ \ac{ENOB}.
\Fig{fig:x_hist} reveals the corresponding histograms of $v_{x_1}, \dots, v_{x_4}$. 
We note that all $v_{x_1}, \dots, v_{x_4}$ voltage swings are well within their guaranteed bounds. However, the histogram of $v_{x_4}$ does not resemble the assumed Gaussian distribution, which warrants the need for the heuristic parameter $\epsilon$ in \Eq{eq:expected_snr}. 


\subsection{Component Variations}\label{sec:component_variations}
\begin{figure}
\centering
\begin{tikzpicture}
    \begin{axis}[
        ybar,
        xlabel={ENOB},
        ylabel={Count},
        grid=both,
        height=0.375\columnwidth,
        width=\columnwidth,
        ymin=0,
        legend style={at={(0.03,0.97)}, anchor=north west},
    ]
    \addplot[bar width=0.04417048881877861, color=blue, fill=blue, opacity=0.6] table [x=bin_center, y=counts, col sep=comma] {data/histogram_ENOB.csv};
    \addlegendentry{$\tau$}
    \addplot[bar width=0.0013805347944728652, color=red, fill=red, opacity=0.5] table [x=bin_center, y=counts, col sep=comma, opacity=0.7] {data/histogram_ENOB_offset.csv};
    \addlegendentry{offset}
    \end{axis}
\end{tikzpicture}
\caption{\label{fig:hist_ENOB}Histogram of estimated \acp{ENOB} for $10^5$ Monte Carlo simulations where the $RC$ values are drawn uniformly at random within $\pm 20 \, \%$ of the nominal values in blue and comparator input offsets drawn uniformly at random within $\pm 20 \, \%$ of $v_\text{max}$ in red. }
\end{figure}
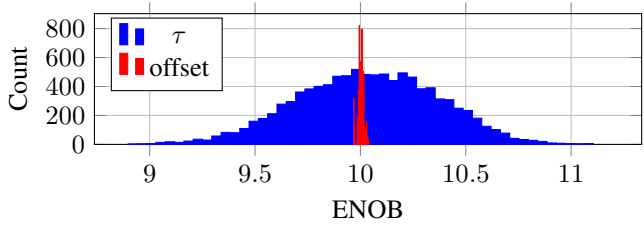

To investigate the sensitivity to component variations, we conducted $10^5$ Monte Carlo simulations, where each individual $RC$ combination in \Fig{fig:RC-chain} was drawn uniformly at random within $\pm 20 \, \%$ of its corresponding nominal value.
Each transient simulation was done as described in \Sec{sec:behavioral_simulations} and the resulting \ac{ENOB} histogram is marked in blue in \Fig{fig:hist_ENOB}. The results show variations within approximately $\pm 1 \, \text{ENOB}$ which is encouraging considering the large range of $RC$ values.

There is no reason to suspect that the operating point of $v_1,\dots,v_4$, would impact the \ac{SNR} performance as long as all voltages remain within valid ranges.
To demonstrate this, we consider the impact of a static offset in the comparator input by some additional $10^5$ Monte Carlo simulations where each comparator input offset is drawn uniformly at random within $\pm 20 \, \% $ of $v_\text{max}$. The resulting \ac{ENOB} histogram is marked in red in \Fig{fig:hist_ENOB}. The results demonstrate the robustness of the RC-chain towards comparator input offsets as the \ac{ENOB} remains within $\pm 0.05 \, \text{ENOB}$. Note that the transient simulations presume that the comparator's speed and sensitivity are unaffected by the shift in operating point. 

\subsection{Digital Loop Delay}
A practical comparator design will incur a delay between a triggering clock edge and a potential change to the inverter output. We denote such a digital delay $\Delta_\text{D}$ and proceed with transient simulations, as in \Sec{sec:behavioral_simulations}, where we also introduce a digital delay up to a full clock period, i.e., $\Delta_\text{D}\in[0,1/f_s]$.
\begin{figure}
\centering
\begin{tikzpicture}
    \begin{axis}[
        ylabel={ENOB},
        xlabel={$\Delta_\text{D}  f_s $},
        grid=both,
        height=0.325\columnwidth,
        width=\columnwidth,
        xmin=0,
        xmax=1,
        legend style={
            at={(0.03,0.03)},   
            anchor=south west,  
        }
    ]
    \addplot[
        thick,
        mark=none] table [x=delay, y=ENOB, col sep=comma] {data/delay_data.csv};   
    \end{axis}
\end{tikzpicture}
\caption{\label{fig:digital_delay}Estimated \ac{ENOB} as a function of digital loop delay $\Delta_D$ from transient simulations with a sinusoidal $v_u$ of amplitude $v_\text{max}$ and angular frequency $\omega_\mathcal{B} / 2$.}
\end{figure}
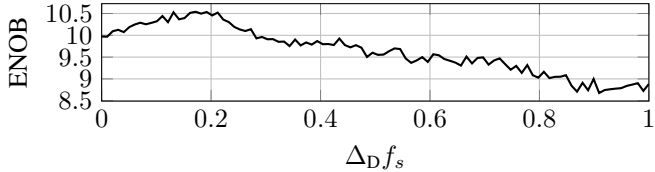
The estimated \ac{ENOB} as a function of $\Delta_\text{D}$ is shown in \Fig{fig:digital_delay}. 
Surprisingly, the results indicate that the highest performance is achieved around $\Delta_\text{D} = 0.2/f_s$ after which the performance gracefully degrades by no more than approximately one \ac{ENOB} for a full clock period delay.

\subsection{Noise Transfer Functions}\label{sec:noise}
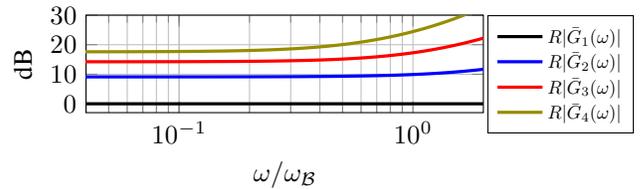
\begin{figure}
\centering
\begin{tikzpicture}
    \begin{semilogxaxis}[
        ylabel={dB},
        xlabel={$\omega / \omega_\mathcal{B}$},
        grid=both,
        xmin=4e-2,
        xmax=2e0,
        ymax=30,
        height=0.325\columnwidth,
        width=0.775\columnwidth,
        legend style={
            at={(1.01,1)},   
            anchor=north west,  
            opacity=1.0,
            legend columns=1,
            inner sep=3pt,       
            nodes={scale=0.7}, 
        }
    ]
    \addplot[
        very thick,
        mark=none] table [x=f_norm, y=G_1_G_1, col sep=comma] {data/tf_data.csv};
    \addlegendentry{$R|\bar{G}_1(\omega)|$}
    
    \addplot[very thick,
        mark=none, color=blue] table [x=f_norm, y=G_1_G_2, col sep=comma] {data/tf_data.csv};
    \addlegendentry{$R|\bar{G}_2(\omega)|$}
    
    \addplot[very thick,
        mark=none, color=red] table [x=f_norm, y=G_1_G_3, col sep=comma] {data/tf_data.csv};
    \addlegendentry{$R|\bar{G}_3(\omega)|$}
    
    \addplot[very thick,
        mark=none, color=olive] table [x=f_norm, y=G_1_G_4, col sep=comma] {data/tf_data.csv};
    \addlegendentry{$R|\bar{G}_4(\omega)|$}
    \end{semilogxaxis}
\end{tikzpicture}
\vspace{-0.435cm}
\caption{\label{fig:transfer_function}Input referred transfer function $|\bar{G}_\ell(\omega)|$ for additive current noise source injected into the $\ell$-th capacitor of \Fig{fig:RC-chain}.
}
\end{figure}

To assess the impact of additive noise sources, such as thermal noise, kickback noise etc., \Fig{fig:transfer_function} presents the input referred transfer function $\bar{G}_\ell(\omega)$ from a test current $i_{C_\ell}$ into each $\ell$-th capacitor.
The transfer functions reveal that later stages of the RC-chain are more susceptible to noise disturbances. Therefore, the rightmost $RC$ nodes in \Fig{fig:RC-chain} will dominate the overall thermal noise sensitivity.  
The input referred thermal noise variance follows as $\bar{v}^2 = 4 k T R \frac{\omega_\mathcal{B}}{2 \pi } \Phi$ [\si{\volt^2}]
where $k$ is the Boltzmann's constant, $T$ is temperature, and $\Phi \eqdef \sum_{\ell=1}^N \frac{\xi_\ell}{ \omega_\mathcal{B} } \int_0^{\omega_{\mathcal{B}}} |\bar{G}_\ell(\omega)|^2 \dd \omega$ (in our example $\Phi \approx 220$).
For $\tau \eqdef RC = 20 \, \si{\nano \second}$, $T=300 \, \si{\kelvin}$, and $v_\text{max}=1 \,\si{\volt}$, the expected squared nominal conversion error equals $\bar{v}^2$ for $C=\num{2.3}$ \si{\pico \farad} and $\{R, R/\kappa_1, \dots, R/\kappa_N\} \approx \{8.9 \text{k},10\text{k}, 58\text{k}, 330\text{k}, 1.8 \text{M} \} \, \si{\ohm}$.


\subsection{Driving the RC-chain ADC}
For the RC-chain \ac{ADC}, due to the switching digital control, the input impedance will not be static but change dynamically over time. 
However, from \Sec{sec:ensuring_reduced_voltages}, it follows that $|v_{x_1}|\leq \delta v_\text{max}$, meaning that $v_{x_1}$ acts as a loose signal ground, making the input impedance fluctuate around $R$. Specifically, excess currents, not larger than $\delta v_\text{max} / R$, will potentially load the input source in addition to the nominal currents $v_\text{in} / R$. 

\section{Outlook}\label{sec:outlook}

The feasibility of the proposed RC-chain \ac{ADC} mainly comes down to the comparator design requirements. 
Specifically, we observe an inherent trade-off between comparator speed and sensitivity, i.e., $\text{OSR} \cdot \omega_\mathcal{B} / \pi$ and $\delta^N v_\text{max}$.
This non-trival tradeoff manifests itself when dealing with the optimal filter length $N$.
On the one hand, a larger $N$ incurs significant in-band filter attenuation, which is a disadvantage, as this increases the sensitivity requirements of all comparators; on the other hand, a larger $N$ simultaneously reduces the relative $\text{OSR}$, effectively reducing the speed requirements of the comparators, see \Fig{fig:snr_surface}.
Thankfully, the overall design, and the comparator in particular, demonstrate robustness towards static mismatch and offsets. However, additional investigations are needed to assess the impact of kickback noise and other dynamic disturbances. 

Another feasibility concern is the resistor \acp{DAC} implementation as these potentially require very large $R$ values, see \Sec{sec:noise}.
Therefore, a mixture of current, resistor, and switched capacitor \acp{DAC} will most likely be considered in an \ac{IC} prototype.

In terms of power consumption, we foresee two potential bottlenecks. Firstly, the aforementioned comparators will consume considerable power in order to manage low voltage swings at high speeds. This applies particularly to the rightmost comparator in \Fig{fig:RC-chain}. The second concern is the thermal noise generated by all the resistors in the filter. Lower resistance values will lower the thermal noise sensitivity; however, at the expense of pulling larger currents out of the resitstor-\acp{DAC} and hence increasing the power consumption. In particular, the leftmost inverter in \Fig{fig:RC-chain} will be the most power hungry.

\section{Conclusion}\label{sec:conclusion}
The continuous-time amplifier-less RC-Chain \ac{ADC} demonstrates how analog-to-digital conversion can be done, in continuous-time, using only passives, comparators, and inverters. 
In addition to a simplistic design, we indicate good robustness properties in respect to component variation, comparator input offset, and modest driving requirements. 
Although further investigation is needed, particularly on the transistor level, our results suggest that this approach could be applicable in a range of \ac{SNR} and \ac{BW} scenarios with special emphasis on technologies where linear amplifiers struggle.
\nopagebreak


\newcommand{\norcas}{IEEE Nordic Circuits and Syst. Conf. (NorCAS)}
\newcommand{\mwscas}{IEEE Int. Midwest Symp. Circuits and Syst. (MWSCAS)}
\newcommand{\iscas}{Proc. IEEE Int. Symp. Circuits Syst. (ISCAS)}
\newcommand{\isscc}{IEEE Int. Solid-State Circuits Conf. (ISSCC)}
\newcommand{\ita}{Information Theory \& Applications Workshop (ITA)}

\newcommand{\tcasi}{IEEE Trans. Circuits Syst. I: Reg. Papers}
\newcommand{\tcasii}{IEEE Trans. Circuits Syst. II: Exp. Briefs}
\newcommand{\procIEEE}{Proceedings of the IEEE}
\newcommand{\vlsi}{IEEE Trans. Very Large Scale Integration (VLSI) Systems}

\newcommand{\jssc}{IEEE J. Solid-State Circuits}
\newcommand{\jestcs}{IEEE J. Emerg. Sel. Topics Circuits Syst.}
\newcommand{\ojsscs}{IEEE Open J. of the Solid-State Circuits Society}
\newcommand{\phd}[3]{#1, \enquote{#2,} Ph.D. dissertation, #3.}

\end{document}